\documentclass[12pt,a4paper]{article}
\usepackage[utf8]{inputenc}
\usepackage[T1]{fontenc}
\usepackage{amsmath}
\usepackage{amsfonts}
\usepackage{amssymb}
\usepackage{graphicx}
\usepackage{hyperref}
\usepackage[numbers,sort&compress]{natbib}
\usepackage{authblk}

\title{Production of spectator neutrons, protons and light fragments on fixed targets at NICA}
\author[1,2]{A.O.~Svetlichnyi}
\author[1,2]{E.Y.~Vasyagina} 
\author[1,2]{S.D.~Savenkov}
\author[1,2]{I.A.~Pshenichnov}
\affil[1]{Institute for Nuclear Research of the Russian Academy of Sciences, Moscow, Russia}
\affil[2]{Moscow Institute of Physics and Technology, Dolgoprudny, Russia}

\newcommand{\firstauthoremail}{aleksandr.svetlichnyy@phystech.edu}
\newcommand{\lastauthoremail}{pshenich@inr.ru}

\begin{document}
 
\date{}

\maketitle

\begin{center}
    \small
    Correspondence: \href{mailto:\firstauthoremail}{\firstauthoremail} (Alexandr Svetlichnyi), 
    \href{mailto:\lastauthoremail}{\lastauthoremail} (Igor Pshenichnov)
\end{center}

\begin{abstract}
The Ultrarelativistic Quantum Molecular Dynamics (UrQMD) model was connected to a set of models called Ablation Monte Carlo (AMC) to identify excited spectator fragments on completion of the UrQMD modeling of nucleus-nucleus collisions and then simulate spectator decays. The UrQMD-AMC approach combines the Minimum Spanning Tree (MST) clustering algorithm, which associates individual nucleons with excited clusters termed prefragments, with decay models of excited nuclei from the Geant4 toolkit. The relevant decay model for the prefragments decay is selected based on their  mass and excitation energy, and includes nuclear evaporation, statistical multifragmentation, and Fermi breakup models. The UrQMD-AMC approach was validated by comparing its results with data on the rapidity and transverse momentum distributions of neutrons and light fragments from 600$A$~MeV Sn + Sn and 10.6$A$~GeV Au + Ag collisions. This approach supplements the evolution of individual nucleons during a nucleus-nucleus collision simulated with UrQMD with nucleon clustering to simulate also the production of spectator nuclear fragments, which are not produced in the UrQMD model alone.
\end{abstract}

\vspace{0.5cm}

\section{Introduction}

The study of hot and dense baryon-rich matter created in nucleus-nucleus collisions is a central point of the Baryonic Matter at Nuclotron (BM@N)~\cite{Kapishin2016} and the Multi-Purpose Detector (MPD)~\cite{MPD2025} experiments at NICA facility. The fixed target BM@N experiment is equipped with a forward hadron calorimeter (FHCal) to measure the energy and spatial distributions of forward nucleons and nuclear fragments emitted in a collision event as well as a beam quartz hodoscope (FQH) and a scintillation wall (ScWall) to identify such protons and fragments~\cite{Volkov2023,Volkov2024}. The signals from these forward detectors can be used to determine the collision centrality and reaction plane~\cite{Guber2020}. In addition, the BM@N experiment will be equipped with the Highly-Granular time-of-flight Neutron Detector (HGND)~\cite{Morozov2025} specifically for detecting neutrons, including forward neutrons. The collider MPD experiment will be equipped with a pair of forward hadronic calorimeters FHCal~\cite{Ivashkin2021,Volkov2021} to detect forward nucleons and fragments on the both sides of the MPD collision point. While some of these fragments are expected to propagate very close to the beam path and thus beyond the FHCal acceptance, a fine transverse segmentation of FHCal will still allow the determination of the reaction plane and the collision centrality~\cite{Ivashkin2021}. 

Most of the nucleons and fragments emitted in the forward direction are remnants of the initial nuclei beyond their hot overlap region, or fireball. These non-participating nucleons are thus termed spectator nucleons and can form spectator fragments. They propagate at small angles with respect to the beam direction and can be registered by the above-mentioned detectors.  In order to calculate the efficiency and acceptance of forward detectors in registering spectators, as well as to simulate the detector responses, a reliable Monte-Carlo model of nucleus-nucleus collision events is required. In addition to describing collision dynamics in the overlap region, the model should accurately describe the production of spectator nucleons and fragments. The procedure of centrality and event plane determination should be based on a reliable model of the production of spectator protons, neutrons as well as spectator nuclear fragments.

The Ultrarelativistic Quantum Molecular Dynamics (UrQMD) model~\cite{Bass1998, Bleicher1999} is widely used to simulate hadronic interactions of nuclei. In particular, the UrQMD results for production of $\pi^+$ and $K^+$ in collisions of 3.2$A$~GeV Ar nuclei with C, Al, Cu, Sn and Pb were compared with the data collected in the BM@N experiment~\cite{BMN2023}. However,  on completion of the UrQMD modeling only hadrons and free nucleons are present, not nuclear fragments. In this work the UrQMD model was connected to a set of models called Ablation Monte Carlo (AMC) to identify excited spectator prefragments on completion of UrQMD modeling of nucleus-nucleus collisions and then simulate prefragment decays into free nucleons and nuclei in their ground state.  In the following such a hybrid approach is denoted as UrQMD-AMC.  The AMC models were employed earlier in the Abrasion-Ablation Monte Carlo for Colliders model with Minimum Spanning Tree clustering (AAMCC-MST) to simulate the evolution of spectator matter in nucleus-nucleus collisions~\cite{Svetlichnyi2020, Nepeivoda2022,Vasyagina2025}.

The production of the spectator fragments and nucleons in collisions of $10.6A$~GeV $^{197}$Au projectiles with heavy (AgBr) nuclei in nuclear emulsion and $600A$ MeV $^{124}$Sn projectiles with a $^{124}$Sn target were calculated with UrQMD-AMC and then compared with experimental data. The pseudorapidity distributions of neutrons, protons, deuterons and $\alpha$-particles in collisions of $3.8A$ GeV $^{124}$Xe projectiles with $^{130}$Xe representing CsI and with $^{184}$W targets were calculated to predict the results of the BM@N and MPD measurements at NICA in the fixed-target mode.

\section{Hybrid UrQMD-AMC model }

In this study, version 3.4 of the UrQMD model~\cite{Bleicher1999,Bass1998} was connected with the MST-clustering and statistical decay models that were previously employed in the AAMCC-MST  model~\cite{Svetlichnyi2020,Nepeivoda2022,Kozyrev2022}. In this hybrid approach, the primary evolution of a nucleus-nucleus collision is modeled using the UrQMD model for the first few hundred fm/$c$. The duration of the evolution time in UrQMD is considered as a parameter of the UrQMD-AMC approach. In this work it is set to 100~fm/$c$ because within the range of initial collision energies at  NICA most of nucleon-nucleon collisions occur earlier than 90 fm/$c$. As proved by UrQMD modeling, the average number of nucleon-nucleon collisions mostly ceases to increase at 100~fm/$c$, and later the number of spectator nucleons remains approximately constant over time. 

Upon completion of the UrQMD modeling,  information about all particles and their collision history is written to a file. The AMC part reads out the file to identify that nucleons which escaped both elastic and inelastic collisions and therefore considered as spectators. The MST-clustering algorithm is then used to identify spectator nucleons associated with spectator prefragments~\cite{Nepeivoda2022}. The MST-clustering is modeled in the coordinate space, and the respective clustering parameter is set to $d = 1.8$~fm~\cite{Nepeivoda2022}.  

In the UrQMD model a nucleus-nucleus collision is simulated by propagating nucleons of both colliding nuclei along their individual classical trajectories. During the UrQMD evolution stochastic elastic and inelastic nucleon-nucleon collisions are simulated. It is a very important question whether prefragment formation can occur earlier, before the time when the last nucleon-nucleon collision takes place. In such a scenario of a dynamic formation process, prefragments propagating in close proximity to each other and to free nucleons may be destroyed and then formed again, and also may exchange nucleons. As pointed out~\cite{Bondorf1997}, the recognition of clusters/prefragments, say, by the Spanning Tree algorithm, becomes very ambiguous in this scenario. The method of Bondorf et al.~\cite{Bondorf1997}, as well as the present implementation of MST yield reliable results only at later times when prefragments have stabilized, meaning most of their constituent nucleons remain bound. The integrity and stability of prefragments has to be confirmed by their spatial separation from each other and by the absence of further nucleon-nucleon collisions at later collision time, after chemical and kinetic freeze-out, typically above 100~fm/$c$. The assumption of the fragment formation upon completion of UrQMD simulations is adopted in other works, in particular, devoted to the formation of hypernuclei, see Buyukcizmeci et al.~\cite{Buyukcizmeci2023} as an example.

As a result of the clustering, the excited prefragments are created, and their excitation energies are calculated using one of three options: (1) the Ericson formula based on the particle-hole model~\cite{Ericson1960}; (2) the parabolic ALADIN approximation~\cite{Botvina1995}, which is tuned to describe fragmentation data for light and heavy nuclear systems; and (3) a combination of the Ericson formula for the peripheral nucleus-nucleus collisions and the ALADIN approximation otherwise. 

The approach (3) was used for all calculations in this work. The decay of the excited prefragments was finally simulated using the Evaporation and Statistical Multifragmentation (SMM) models from the Geant4 toolkit \cite{Allison2016}, version 10.4, and the Fermi Break-up (FBU) model from the Geant4 toolkit, version 9.6. All three decay models are involved in the decay of the excited nuclei: the FBU model is used to simulate the decay of nuclei with $A < 20$ and $Z < 9$; the evaporation model is used to simulate the decay of moderately excited heavier nuclei; and the SMM model is used for heavier nuclei with excitation energies higher than 3 MeV/nucleon. The Coulomb repulsion effects specifically between protons of spectator matter are included in our modelling with the AMC, see Vasyagina et al.~\cite{Vasyagina2025} for  detail.

UrQMD provides two options for modeling collision events~\cite{Bass1998}. The first option (Skyrme mode) implements the density-dependent nucleon-nucleon forces. The second option (cascade mode) implements the UrQMD modeling without accounting for the density-dependent nucleon-nucleon forces. It is worth noting that the simulations in the cascade mode are significantly faster than those in the Skyrme mode.

\section{Validation of UrQMD-AMC}

\begin{figure}[htb!]
\begin{minipage}{0.49\linewidth}
    \includegraphics[width=1.1\linewidth]{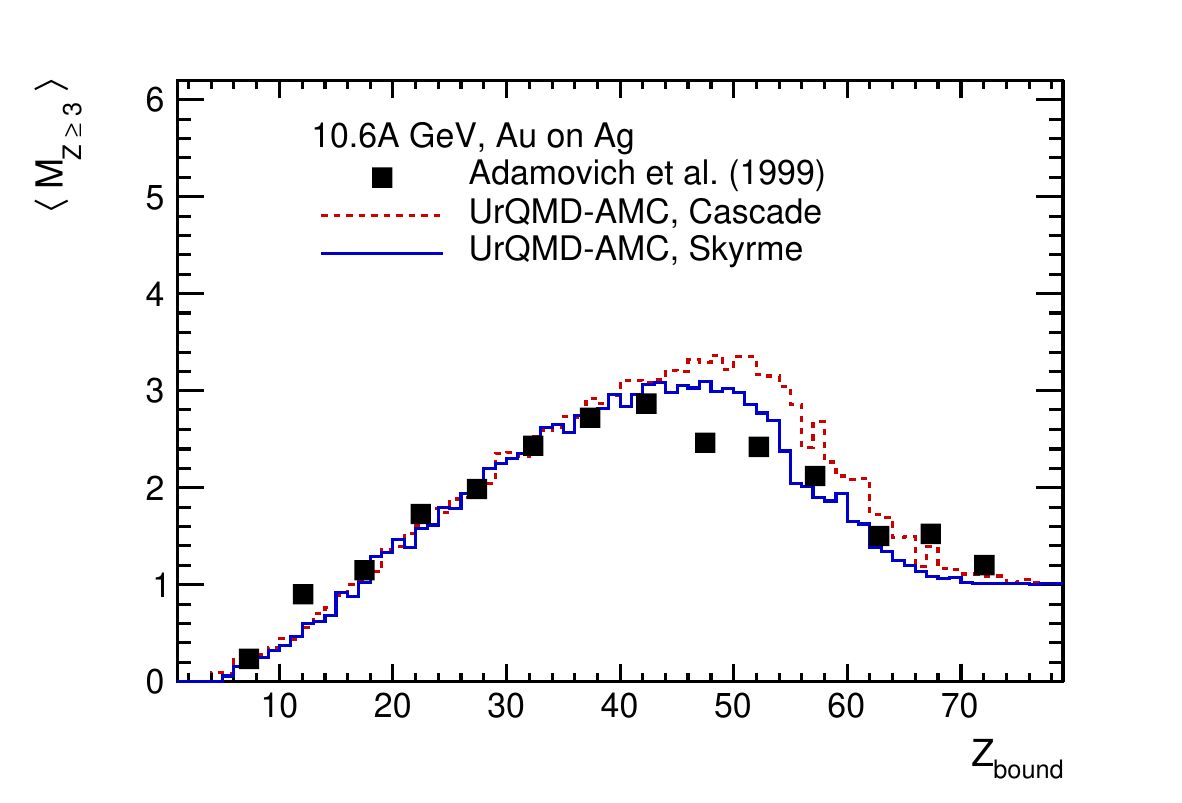}
\end{minipage}
\begin{minipage}{0.49\linewidth}
    \includegraphics[width=1.1\linewidth]{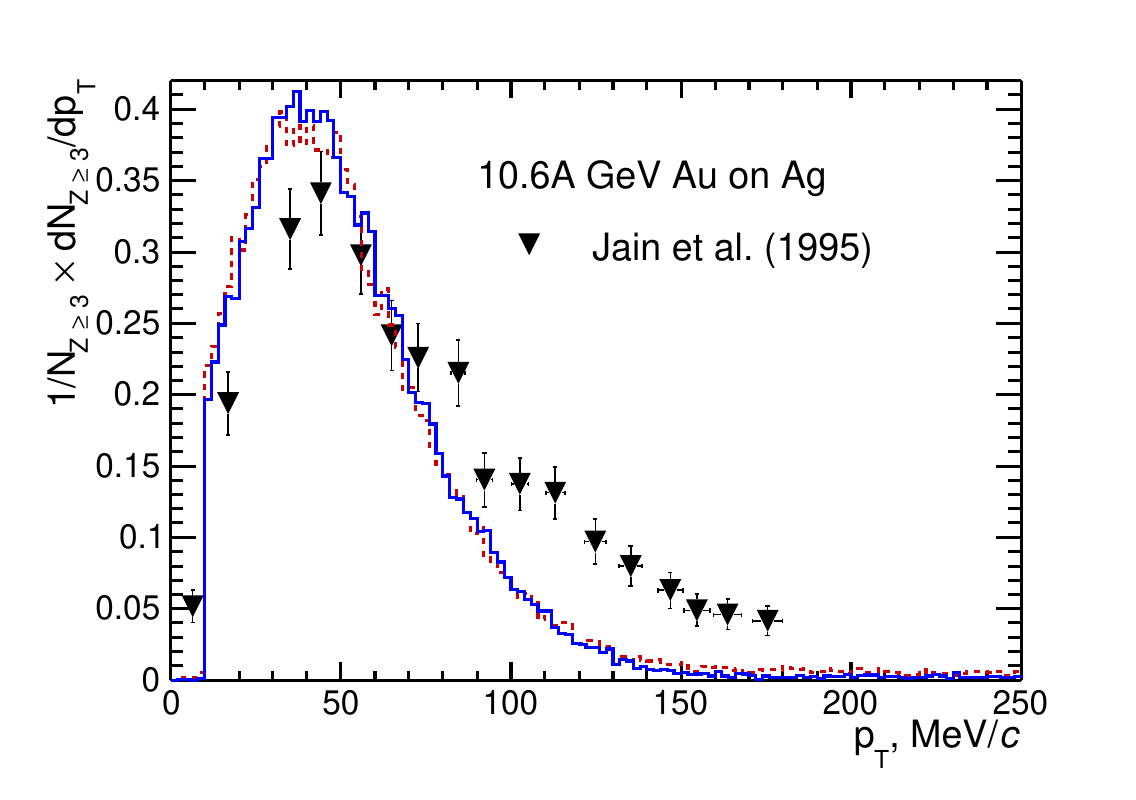}
    \end{minipage}
    \caption{Production of fragments with $Z \geq 3$  in the interactions of $10.6A$~GeV $^{197}$Au with Ag nuclei of nuclear emulsion calculated with UrQMD-AMC in the cascade mode (red dashed-line histograms) and with the Skyrme forces (blue solid-line histograms). Left panel: the average multiplicity of $Z \geq 3$ fragments as a function of the total charge bound in fragments $Z_{\rm bound}$ (see text for details), the black squares represent the experimental data by the EMU01 Collaboration~\cite{Adamovich1999}. Right panel: the transverse momentum distribution of $Z\geq 3$ fragments, the triangles represent the  data  by P. Jain {\em et al.}~\cite{Jain1995}.}
    \label{fig:frags}
\end{figure}

The collisions of 10.6$A$~GeV $^{197}$Au projectiles with Ag nuclei in nuclear emulsion were modeled with UrQMD-AMC. The results on the production of forward charged fragments  were compared with the data on fragmentation of 10.6$A$~GeV $^{197}$Au in collisions with heavy nuclei (ArBr) in nuclear emulsion~\cite{Adamovich1999,Jain1995}. In Fig.~\ref{fig:frags} (left panel) the calculated average multiplicity  $\langle M_{Z\geq 3} \rangle$ of fragments with $Z \geq 3$ is presented as a function of the total charge $Z_{\rm bound}$ bound in fragments with $Z \geq 2$:
\begin{equation}
Z_{\rm bound} = \sum\limits_{Z \geq 2} Z_f \ . 
\nonumber
\end{equation}

The calculations were performed using two modes of the UrQMD model: the cascade mode and the Skyrme model. Silver nuclei were used as the target nuclei instead of ArBr to simplify the calculations. As can be seen in Fig.~\ref{fig:frags}, the data on $\langle M_{Z\geq 3} \rangle$\cite{Adamovich1999} for $Z_{\rm bound} < 40$ are described very well by UrQMD-AMC in both modes of calculation. The specific global evolution of the measured $\langle M_{Z\geq 3} \rangle$ with the increase of $Z_{\rm bound}$ is a well-known phenomenon, understood as the rise and fall of multifragmentation of spectator matter with variation of collision centrality~\cite{Ogilvie1991}. While the general trend of $\langle M_{Z\geq 3} \rangle$ is well reproduced by both calculation  options, the results with the Skyrme option for $Z_{\rm bound} > 40$ are closer to the data.  However, the model overestimates $\langle M_{Z\geq 3} \rangle$  for $ 45 < Z_{\rm bound} < 55 $ by 20--25\% in both modes. 

The UrQMD-AMC was validated also with the transverse momentum $p_{\rm T}$ distribution of $Z \geq 3$ fragments measured by P.~Jain {\em et al.}~\cite{Jain1995}. Calculated $p_{\rm T}$ distributions are compared with the data in Fig.~\ref{fig:frags} (right panel). While the position of the maximum of the distribution and its absolute value is reproduced by modeling, the calculations with both UrQMD-AMC options noticeably underestimate the production of $Z \geq 3$ fragments above 75~MeV/$c$. 

\begin{figure}[htb!]
\begin{minipage}{0.49\linewidth}    
    \centering
    \includegraphics[width=1.1\linewidth]{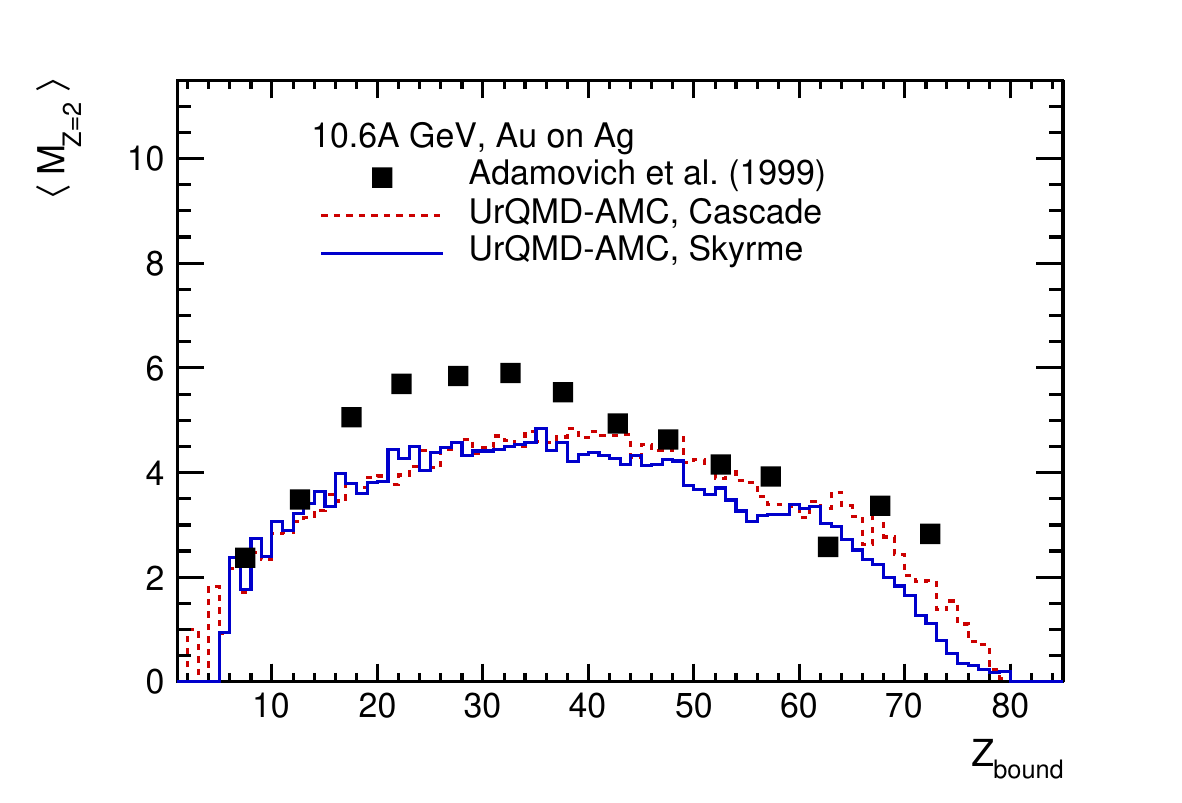}
\end{minipage}
\begin{minipage}{0.49\linewidth}
    \centering
    \includegraphics[width=1.1\linewidth]{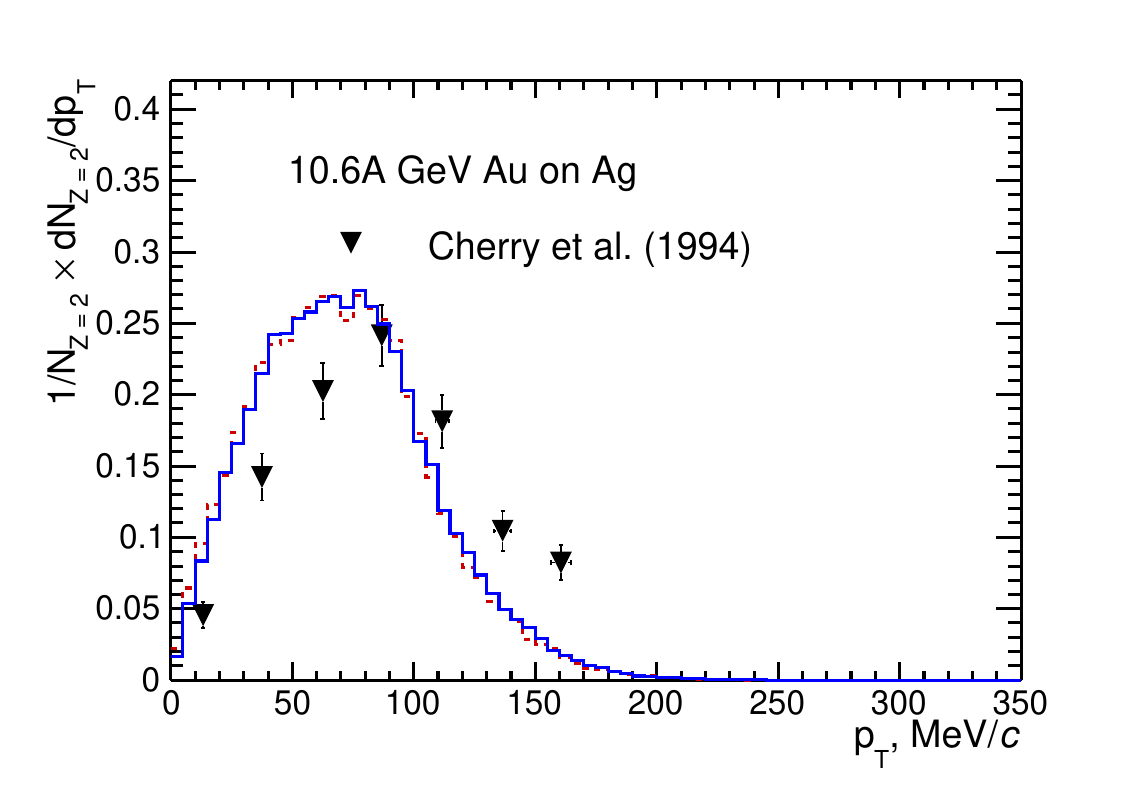}
\end{minipage}
    \caption{Production of helium ($Z=2$) fragments in the interactions of $10.6A$~GeV $^{197}$Au with Ag nuclei of nuclear emulsion calculated with UrQMD-AMC in the cascade mode (red dashed-line histograms) and with Skyrme forces (blue solid-line histograms). Left panel: the average multiplicity of $Z=2$ fragments as a function of the total charge bound in fragments $Z_{\rm bound}$ (see text for details), the black squares represent the experimental data by the EMU01 Collaboration~\cite{Adamovich1999}. Right panel: the transverse momentum distributions of $Z=2$ fragments. The black triangles represent the experimental data by the KLMM~\cite{Jain1995} collaboration.}
    \label{fig:alphas}
\end{figure}

 In Fig.~\ref{fig:alphas} (left panel) the calculated average multiplicity  $\langle M_{Z=2} \rangle$  of helium fragments is presented as a function of $Z_{\rm bound}$ for the same collision system as in Fig.~\ref{fig:frags}. Calculations with both UrQMD-AMC options well describe the data~\cite{Adamovich1999} for $Z_{\rm bound} < 15$ and $Z_{\rm bound} > 40$. However, $\langle M_{Z=2} \rangle$ is underestimated  for $15 < Z_{\rm bound} < 40$ by 20--25\%. The inclusion of the Skyrme forces in calculations does not significantly change $\langle M_{Z=2} \rangle$. This indicates that helium fragments are formed mostly at later stages of the collision, via their evaporation from spectator prefragments or in multifragment decays of spectator matter. 
 
 The transverse momentum $p_{\rm T}$ distribution for helium fragments is presented in Fig.~\ref{fig:alphas} (right panel).  Calculations with the UrQMD-AMC with both options provide a reasonable agreement with the data~\cite{Cherry1994} for  intermediate $p_{\rm T}$. However, as for $Z\geq 3$ fragments, see Fig.~\ref{fig:frags}, the high-$p_{\rm T}$ tail for $Z=2$ fragments is underestimated by UrQMD-AMC. 

 Possible reasons for such an underestimation can be seen. By definition, see Kuttan et al.~\cite{Kuttan2023}, those nucleons which escaped any kind of collision during the UrQMD evolution, either elastic or inelastic, are considered as spectators. Any nucleon which participated at least in one elastic or inelastic interaction is then considered as a participant nucleon. While the present work is focused on spectator matter, these definitions are also adopted for the sake of consistency. In the present approach prefragments are built of spectator nucleons mostly propagating forward with the beam rapidity and very low $p_{\rm T}$.  On one hand, those nucleons which undergo only elastic scattering can also form  $Z=2$ and $Z\ge 3$ prefragments, possibly with a high total $p_{\rm T}$, and hence improve the agreement with data in Figs. \ref{fig:frags} and ~\ref{fig:alphas}. On the other hand, as proved by Kuttan et al.~\cite{Kuttan2023}, attributing elastically scattered nucleons to spectators underestimates the number of participants in UrQMD and makes this model equivalent to the Glauber model which considers only inelastic nucleon-nucleon collisions. The latter approach is justified at high energies of RHIC and LHC where a low ($<15$\%) contribution of the elastic nucleon-nucleon cross section to the total cross section is reported. In contrast, the elastic and inelastic cross nucleon-nucleon cross sections are comparable at NICA energies, and elastic scattering can not be neglected. In future studies we plan to implement clusterization also for participant nucleons. This will require the implementation of clusterization algorithms both in the coordinate and momentum spaces. 

Since the collisions of $^{124}$Xe nuclei with a CsI target as a substitute for a $^{130}$Xe target, were studied in the BM@N experiment~\cite{Zubankov2025}, it is reasonable to validate UrQMD-AMC with measurements performed for collisions of nuclei of similar mass. Recent measurements of neutron emission by $600A$~MeV $^{124}$Sn projectiles in collisions with a $^{124}$Sn target~\cite{Pawlowski2023} provide valuable data for validating UrQMD-AMC.

\begin{figure}[htb!]
\begin{minipage}{0.49\linewidth}    
    \centering
    \includegraphics[width=1.1\linewidth]{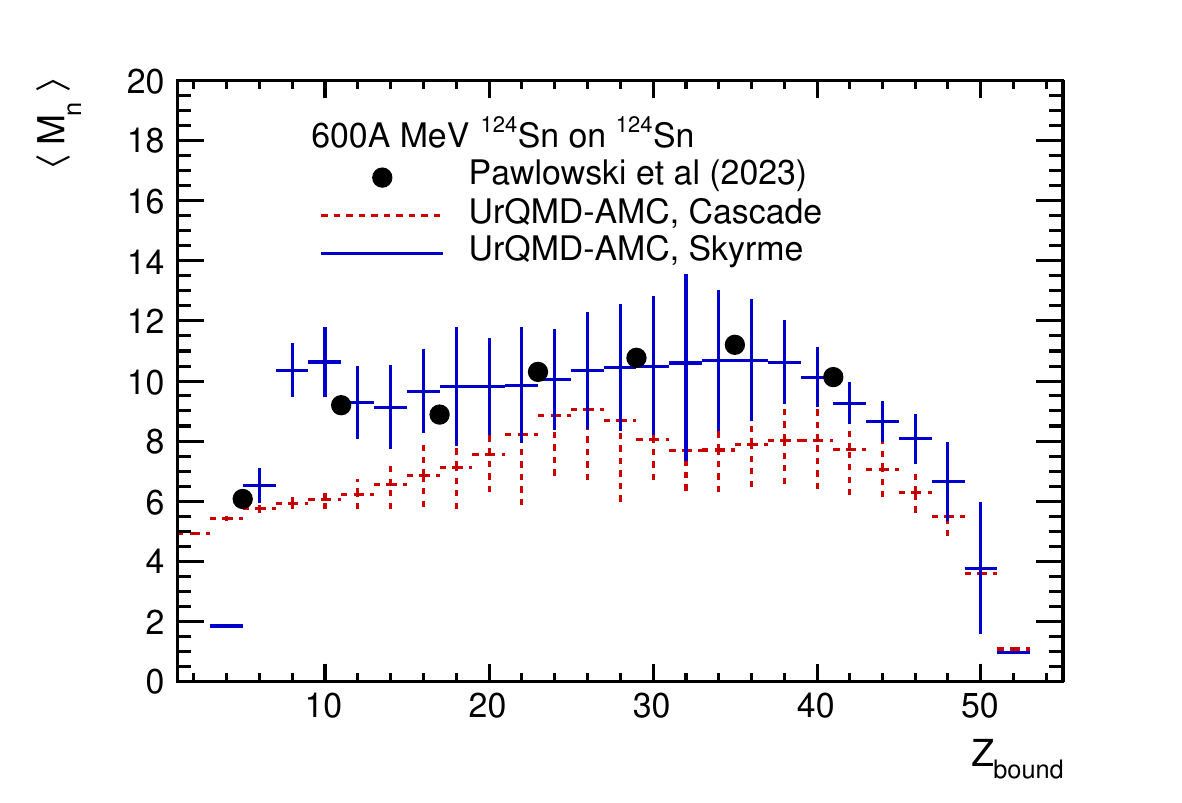}
\end{minipage}
\begin{minipage}{0.49\linewidth}
    \centering
    \includegraphics[width=1.1\linewidth]{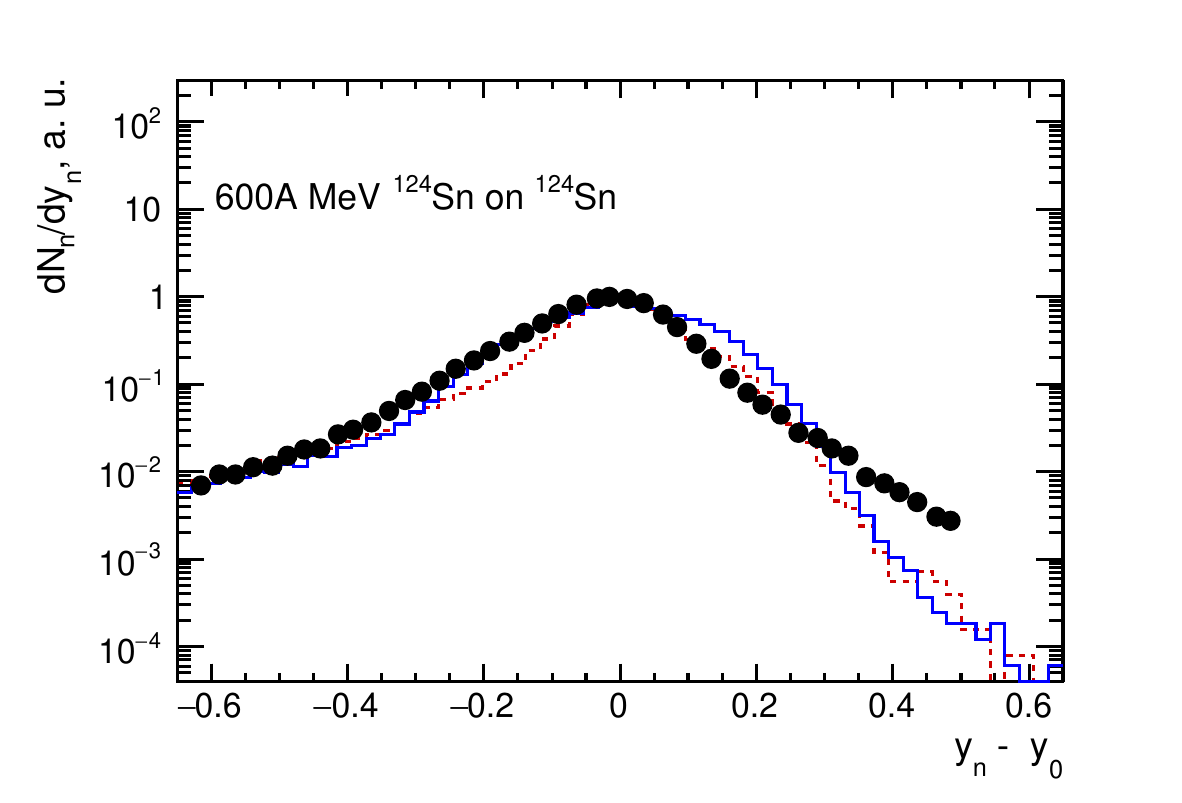}
\end{minipage}
    \caption{Average multiplicity of neutrons as a function of $Z_{\rm bound}$ with its standard deviation represented by error bars calculated for collisions of $600A$~MeV $^{124}$Sn with $^{124}$Sn (left panel). The rapidity distribution of neutrons in the rest frame of the emitting nucleus calculated for the same collision system (right panel).  The circles represent measurements~\cite{Pawlowski2023}, other notations are the same as in Fig~\ref{fig:frags}.}
    \label{fig:neutrons}
\end{figure}

In Fig.~\ref{fig:neutrons} (left panel) the calculated average multiplicity of projectile spectator neutrons $\langle M_{\rm n}\rangle$ is presented as a function of $Z_{\rm bound}$. In addition, the standard deviations of the calculated $\langle M_{\rm n}\rangle$ are represented by error bars. The UrQMD-AMC calculation with the Skyrme forces well describes the data~\cite{Pawlowski2023}. However, the calculation in the cascade mode systematically underestimates the measured $\langle M_{\rm n}\rangle$. This indicates that neutrons are produced mostly at the first UrQMD stage of modeling. 

In Fig.~\ref{fig:neutrons} (right panel) the rapidity distribution of projectile spectator neutrons in the rest frame of the emitting nucleus is presented. The beam rapidity is denoted as $y_0$, and it is subtracted from neutron rapidity $y_{\rm n}$ measured in the laboratory frame. The UrQMD-AMC with Skyrme forces better describes the data~\cite{Pawlowski2023} for neutrons with $y_{\rm n}<y_0$. However, in both calculation modes the yields of very forward neutrons with $y_{\rm n}-y_0>0.3$ are underestimated.

As can be seen,  a better description of forward neutron production in $^{124}$Sn--$^{124}$Sn collisions is obtained with accounting for the Skyrme forces. In contrast, the calculated multiplicities of $Z=2$ and $Z\geq 3$ fragments produced Au--Ag collisions are less sensitive to the inclusion of the Skyrme forces.

\section{Production of neutrons, protons, deuterons and \texorpdfstring{$\alpha$}{alpha}-particles in collisions of \texorpdfstring{$^{124}$Xe}{124} with fixed targets}

After validating the UrQMD-AMC approach with available experimental data, the pseudorapidity distributions of nucleons and light fragments to be measured in future experiments at NICA can be calculated. During the recent runs of the BM@N experiment, the collisions of $3.8A$~GeV $^{124}$Xe nuclei with the CsI target equivalent to a $^{130}$Xe target were studied~\cite{Zubankov2025}. It is expected that the physics program at NICA collider will start with collisions of one of the $^{124}$Xe beams in the collider ring with a thin $^{184}$W target.
The electromagnetic dissociation of $^{124}$Xe nuclei on fixed targets, $^{130}$Xe and $^{184}$W, respectively, in the BM@N and MPD experiments has been recently considered~\cite{Pshenichnov2024}. In the present work the results for hadronic fragmentation of $^{124}$Xe are given. 

\begin{figure}[htb!]
\begin{minipage}{0.49\linewidth}    
    \centering
    \includegraphics[width=1.1\linewidth]{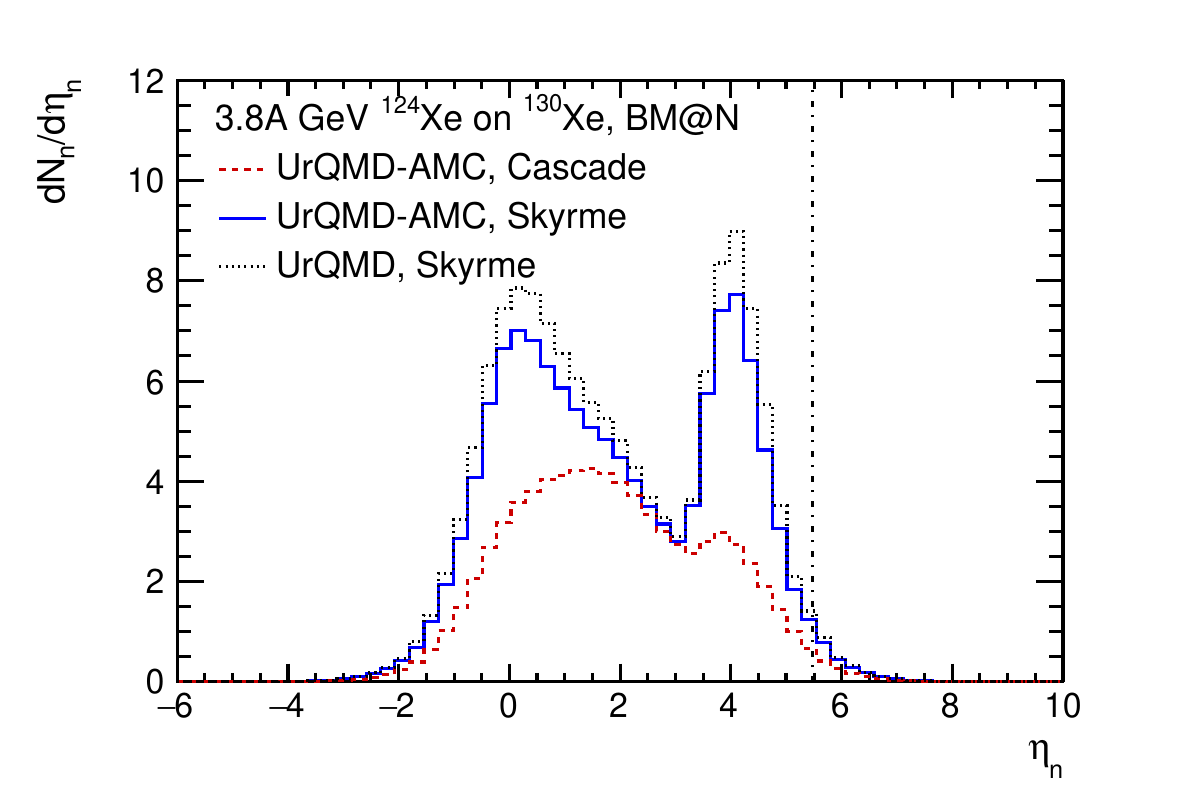}
\end{minipage}
\begin{minipage}{0.49\linewidth}
    \centering
    \includegraphics[width=1.1\linewidth]{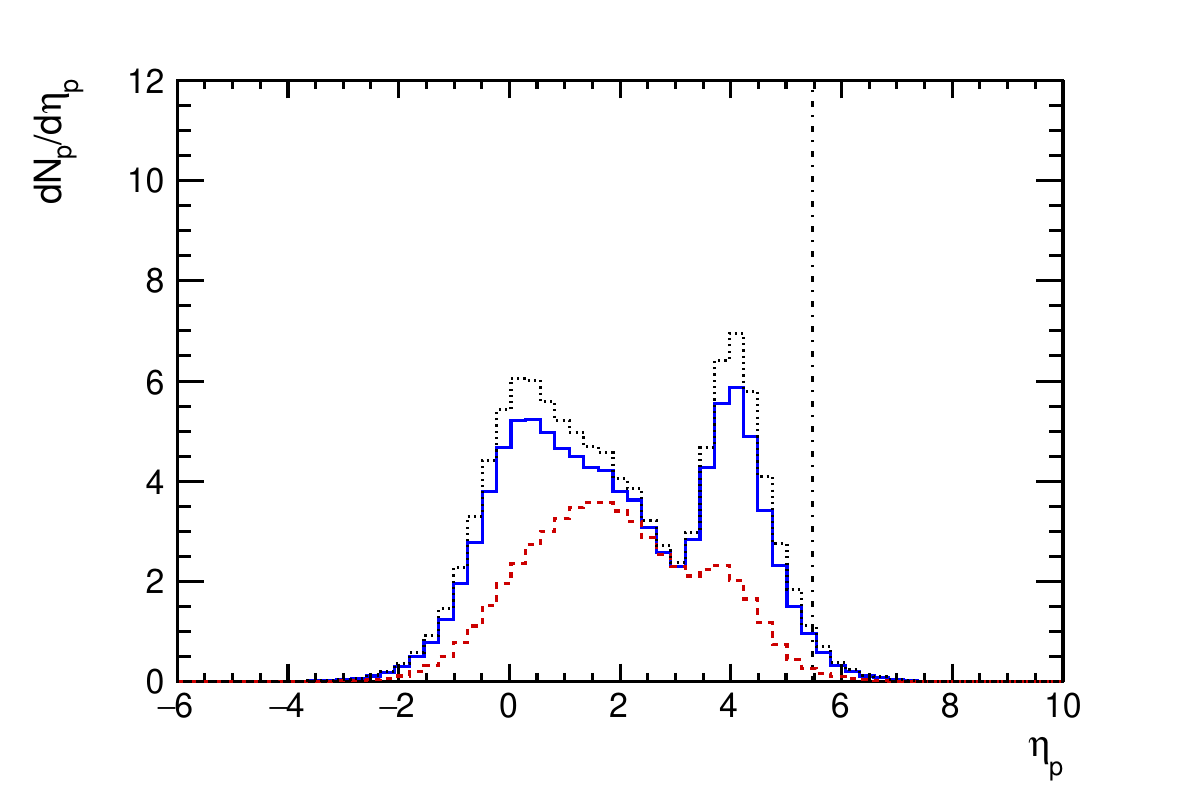}
\end{minipage}

\begin{minipage}{0.49\linewidth}    
    \centering
    \includegraphics[width=1.1\linewidth]{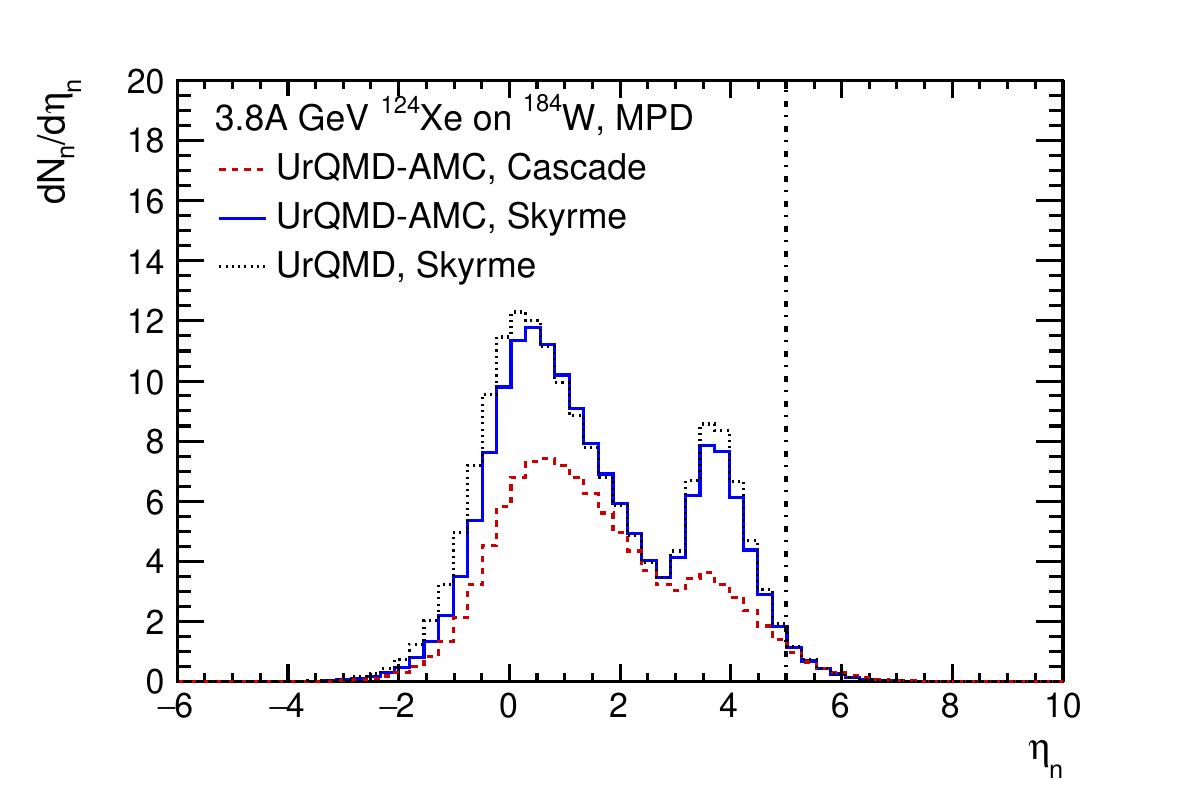}
\end{minipage}
\begin{minipage}{0.49\linewidth}
    \centering
    \includegraphics[width=1.1\linewidth]{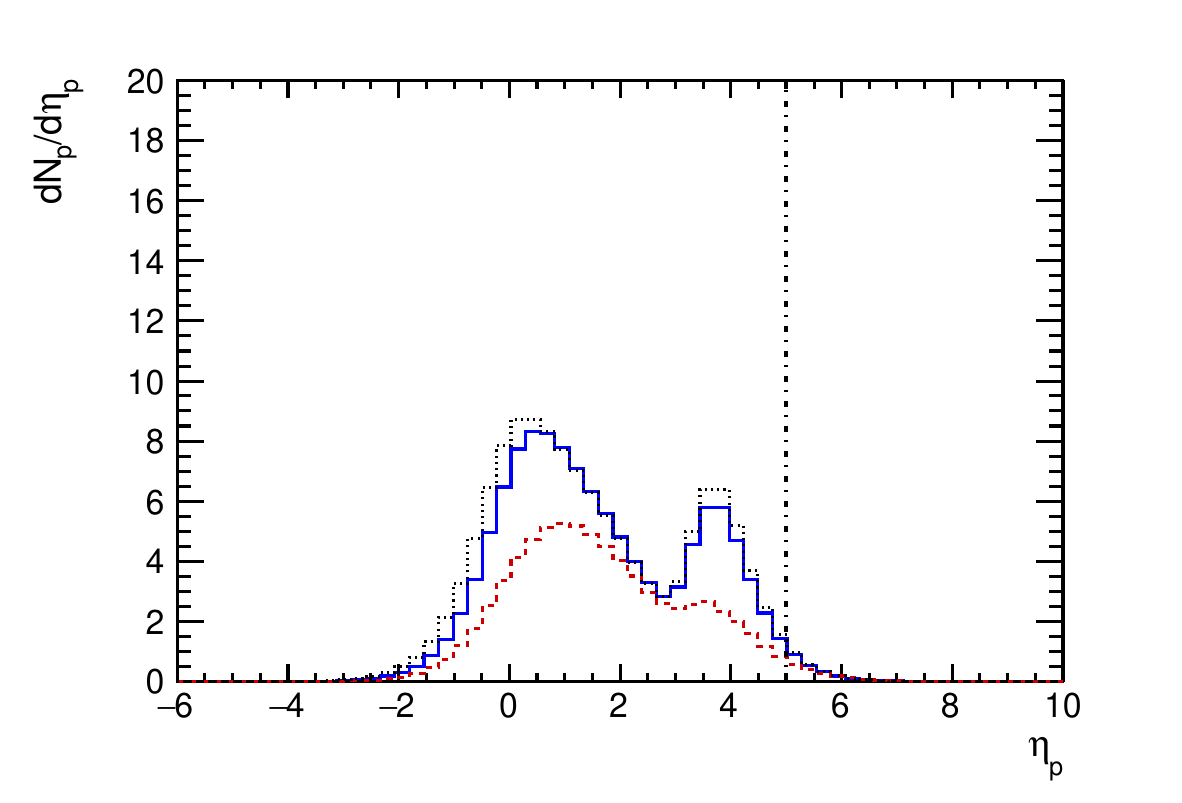}
\end{minipage}

    \caption{Calculated pseudorapidity distributions of neutrons (left) and protons (right) from collisions of $3.8A$~GeV $^{124}$Xe with a $^{130}$Xe target in the BM@N experiment (top), and with a $^{184}$W target in the MPD experiment (bottom). Dotted vertical lines mark the upper boundary of the pseudorapidity interval covered by forward calorimeters in the experiments.}
    \label{fig:n_p_eta}
\end{figure}

The pseudorapidity distributions ${\rm d}N/{\rm d}\eta$ calculated with UrQMD-AMC for neutrons (left) and protons (right) produced in fragmentation of 3.8$A$~GeV $^{124}$Xe in the BM@N (top) and MPD (bottom) experiments are presented in Fig.~\ref{fig:n_p_eta}. The distributions demonstrate two distinct peaks corresponding to the emission from target and projectile nuclei, respectively. The results of UrQMD modeling with the Skyrme forces, but without AMC, are also shown in Fig.~\ref{fig:n_p_eta}. As can be seen from the comparison of UrQMD-AMC and UrQMD results, some 10\% of free nucleons produced in minimum bias events  remain bound in fragments as soon as AMC is activated. The heights of the peaks corresponding to nucleons emitted from $^{124}$Xe projectiles are approximately equal in collisions with $^{130}$Xe and $^{184}$W targets. However, as expected, the number of nucleons with $\eta\approx 0$ emitted from more heavy $^{184}$W target nuclei with respect to $^{130}$Xe is noticeably higher. The UrQMD-AMC results for  ${\rm d}N/{\rm d}\eta$ obtained with the cascade and Skyrme mode are different. Much less free nucleons are produced in the cascade mode which also predicts more intense stopping as the respective peaks at $\eta=0$ and the projectile rapidity are shifted more towards each other in comparison to modeling in the Skyrme mode.  

\begin{figure}[ht!]
\begin{minipage}{0.49\linewidth}    
    \centering
    \includegraphics[width=1.1\linewidth]{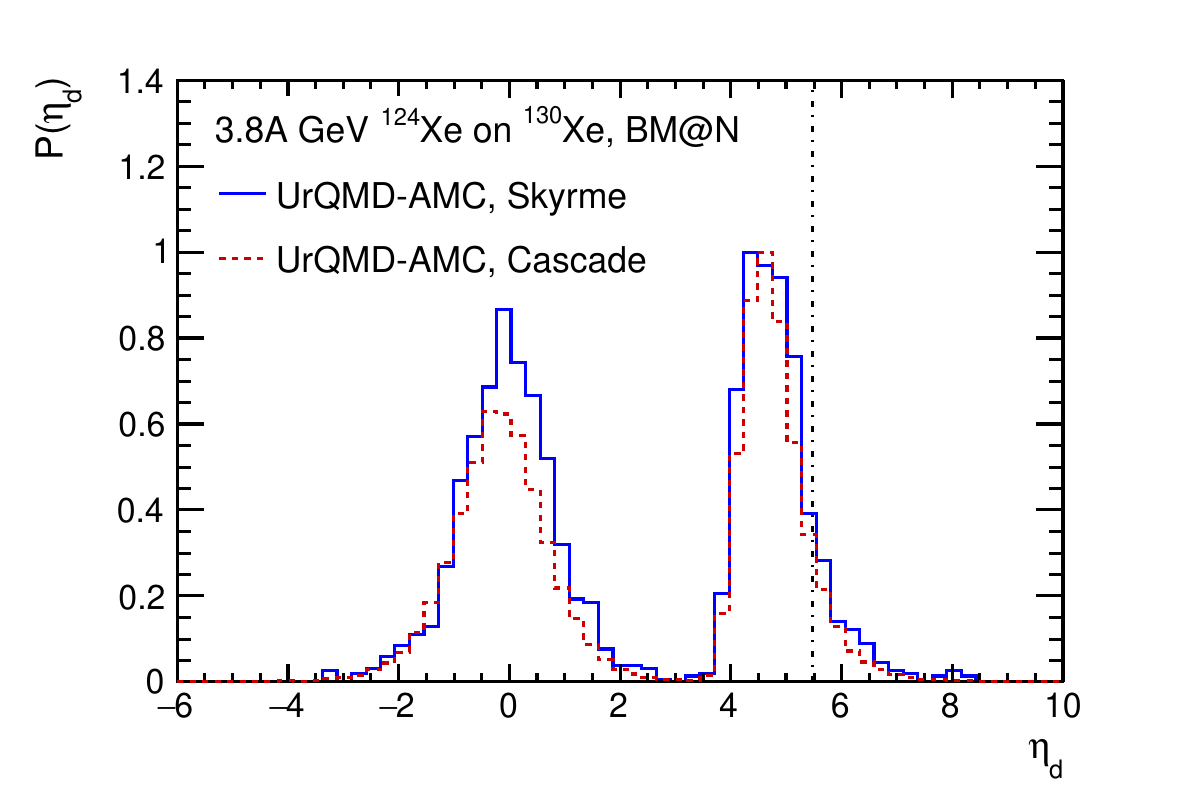}
\end{minipage}
\begin{minipage}{0.49\linewidth}
    \centering
    \includegraphics[width=1.1\linewidth]{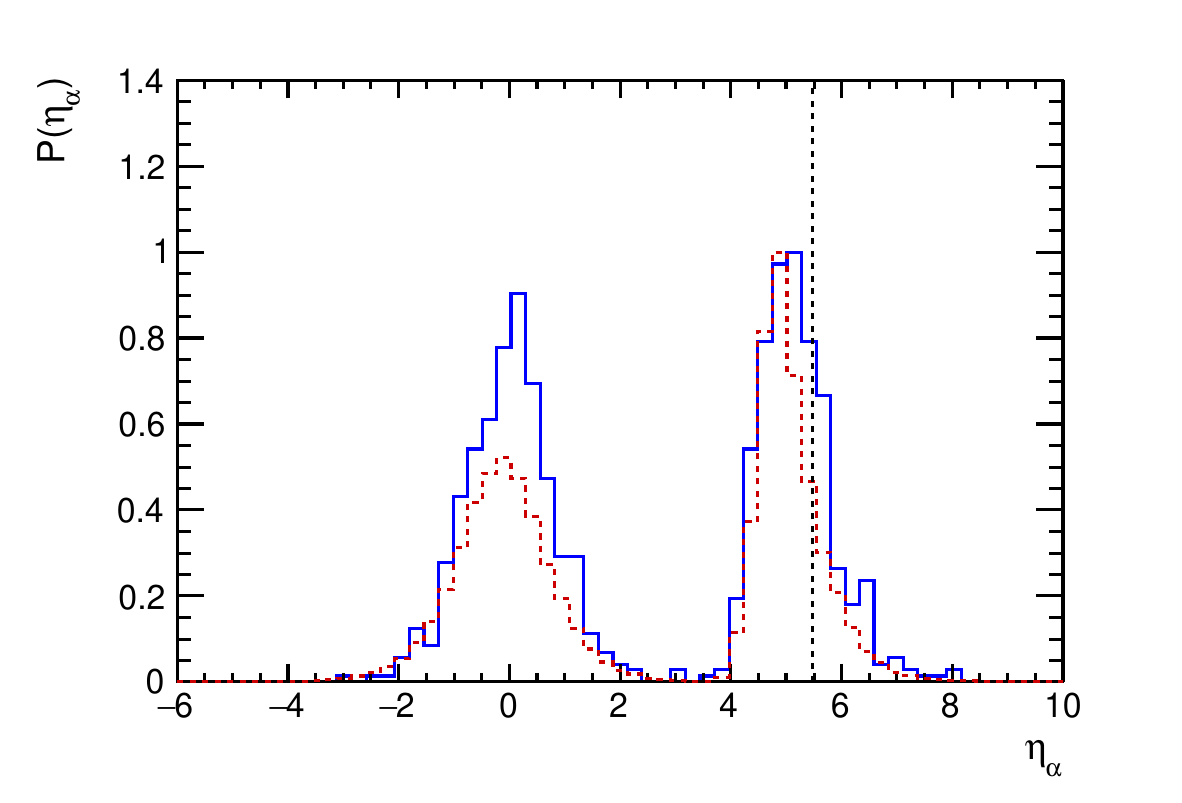}   
\end{minipage}

\begin{minipage}{0.49\linewidth}    
    \centering
    \includegraphics[width=1.1\linewidth]{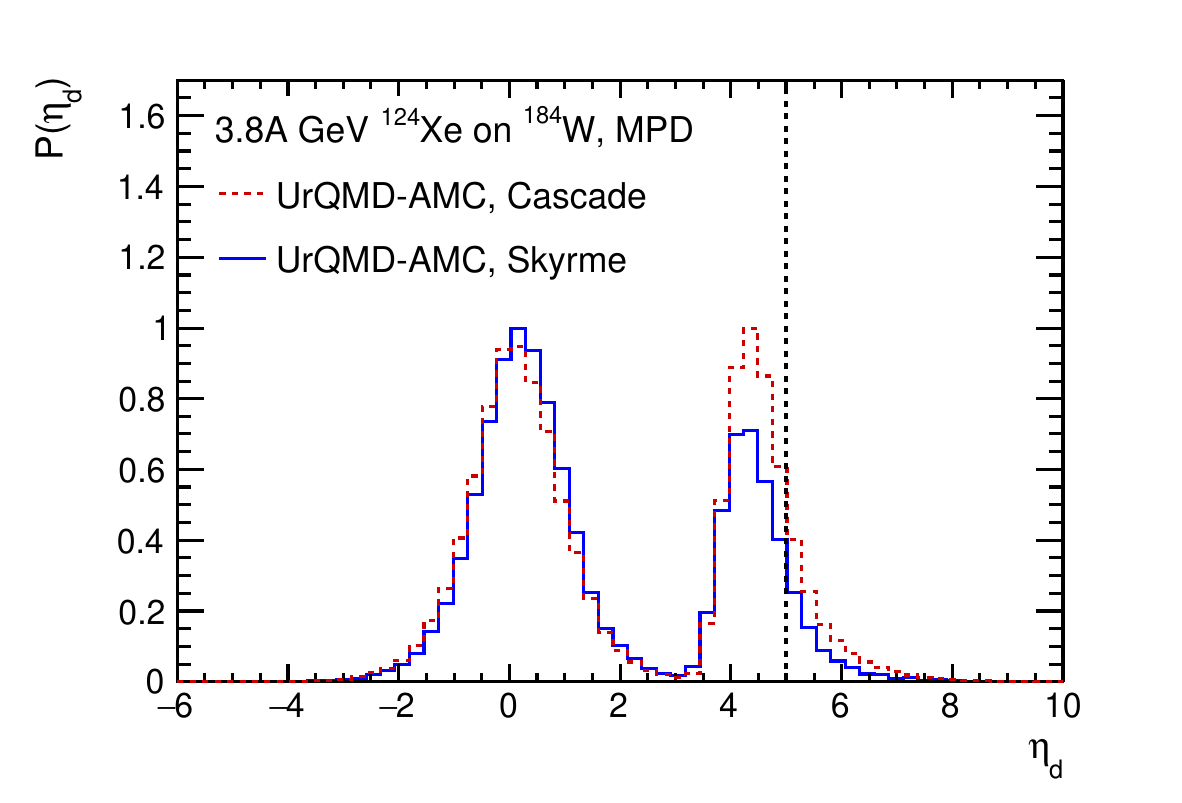}
\end{minipage}
\begin{minipage}{0.49\linewidth}
    \centering
    \includegraphics[width=1.1\linewidth]{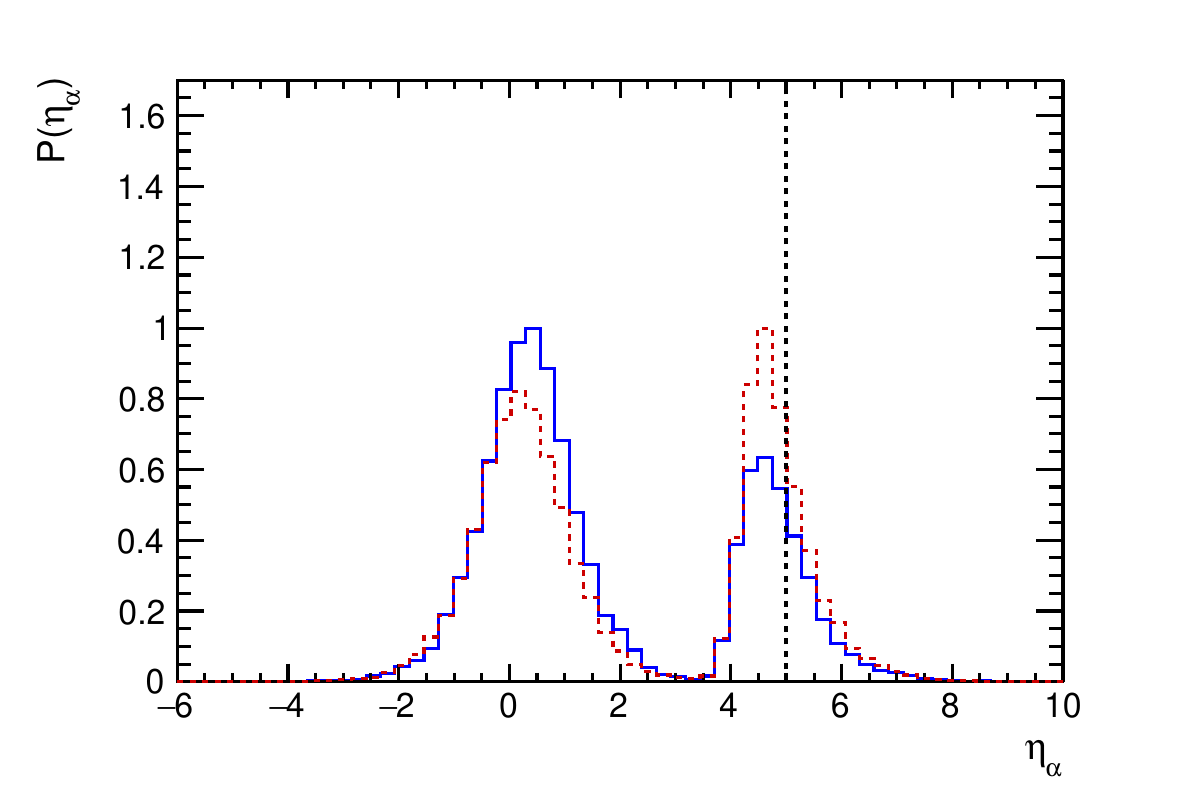}
\end{minipage}

    \caption{Same as in Fig.~\ref{fig:n_p_eta}, but for deuterons (left) and $\alpha$-particles (right).}
    \label{fig:a_d_eta}
\end{figure}

The pseudorapidity distributions ${\rm d}N/{\rm d}\eta$ calculated with UrQMD-AMC for deuterons (left) and $\alpha$-particles (right) produced in fragmentation of 3.8$A$~GeV $^{124}$Xe in the BM@N (top) and MPD (bottom) experiments are presented in Fig.~\ref{fig:a_d_eta}.  It is interesting to note that in the cascade mode of UrQMD-AMC 
less deuterons and $\alpha$-particles are produced in  $^{124}$Xe--$^{130}$Xe collisions in comparison to the Skyrme mode calculations. In contrast, the modeling of $^{124}$Xe--$^{184}$W in the cascade mode provides more deuterons and $\alpha$-particles emitted from $^{124}$Xe projectiles in comparison to the Skyrme mode. This indicates the target-size dependence of the modeling results obtained in the cascade and Skyrme modes and also demonstrates typical systematic uncertainties of the modeling which are ineviably propagated to the systematic uncertainties of acceptance and efficiency calculations performed for the BM@N and MPD experiments.  
 
The dotted vertical lines in Figs.~\ref{fig:n_p_eta} and~\ref{fig:a_d_eta} correspond to the radius of the beam hole and thus approximately mark the upper limit of the pseudorapidity acceptance interval of the forward calorimeters of BM@N and MPD. 
As seen, a small number of nucleons are not intercepted by the forward calorimeters, while about $1/3$ of $\alpha$-particles are expected to propagate through the beam hole and remain undetected.

\section{Conclusion}

In our hybrid UrQMD-AMC approach, the multiparticle evolution of individual nucleons simulated with UrQMD~\cite{Bass1998,Bleicher1999} is supplemented by their clustering to model the production of spectator nuclear fragments. In the present work the UrQMD-AMC approach was validated by calculating the production of $Z\geq 3$ and $Z=2$ fragments in $^{197}$Au--Ag collisions and neutrons in $^{124}$Sn--$^{124}$Sn collisions and comparing the results with the respective experimental data~\cite{Adamovich1999,Jain1995,Cherry1994,Pawlowski2023}. Using the UrQMD model with the Skyrme forces yields better agreement with the data than the cascade mode. However, the average fragment multiplicities for semi-central or semi-peripheral collisions were underestimated by the calculations in both modes by 20--25\%. The UrQMD-AMC approach also underestimates the production of fragments with high transverse momenta, possibly due to neglecting the clusterization of elastically scattered nucleons. Nevertheless, the production of neutrons in $^{124}$Sn--$^{124}$Sn collisions is well described.  

The UrQMD-AMC modeling of $^{124}$Xe--$^{130}$Xe and $^{124}$Xe--$^{184}$W collisions to be studied at NICA demonstrated a noticeable difference between the calculated pseudorapidity distributions of nucleons obtained in the cascade and Skyrme modes of UrQMD. However, similar distributions calculated for deuterons and $\alpha$-particles were found to be less sensitive to the choice of the UrQMD mode. As found, most of forward neutrons and protons are emitted within the acceptance of the forward calorimeters of the BM@N and MPD experiments with fixed targets. However, approximately 30\% of spectator $\alpha$-particles are expected to remain undetected.

In summary, this work demonstrated the deviations of the UrQMD-AMC results from the available experimental data and the differences in the calculation results obtained with different UrQMD modes. These differences characterize the range of systematic uncertainties of future NICA measurements stemming from the calculated corrections for detector efficiency and acceptance that will be applied to raw experimental data.

\section*{Acknowledgements}
The work supported by the Ministry of Science and Higher Education of the Russian Federation, Project FFWS-2024-0003.

\bibliographystyle{elsarticle-num}

\bibliography{urqmd-amc}

\end{document}